\documentclass[]{raa}
\usepackage{graphicx}
\usepackage{times}             
\usepackage{amssymb}
\usepackage{amsmath}
\usepackage{mathrsfs}
\usepackage{float}
\usepackage{natbib}
\usepackage{booktabs}
\usepackage{hyperref}
\usepackage{xspace}

\begin{document}

   \title{The Timing Residual Patterns Due to Pulsar Acceleration}

   \setcounter{page}{1}

   \author{Yi Xie
      \inst{1}
   \and
      Li-Chun Wang \inst{1,2} }


\institute{School of Science, Jimei University, Xiamen 361021, Fujian Province, China; {\it xieyi@jmu.edu.cn}\\
\and
Physics Experiment Center, Jimei University, Xiamen 361021, Fujian Province, China; {\it wanglc@jmu.edu.cn}}


\vs \no
   {\small Received 2020 Month Day; accepted 20XX Month Day}

\abstract{The form of timing residuals due to errors in pulsar spin period $P$ and its derivative $\dot{P}$, in positions, as well as in proper motions, have been well presented for decades in the literature. However, the residual patterns due to errors in the pulsar acceleration have not been reported previously, while a pulsar in the galaxy or a globular cluster (GC) will be unavoidably accelerated. The coupling effect of the pulsar transverse acceleration and the R$\rm{\ddot{o}}$mer delay on timing residuals are simulated in this work. The results show that the residual due to the effect can be identified by the oscillation envelopes of the residuals. It is also shown that the amplitude of the residual due to the effect is usually relatively small, however, it may probably be observable for pulsars distributing in the vicinity of the core of a nearby GC.
}

   \authorrunning{Yi Xie \& Li-Chun Wang}            
   \titlerunning{Residual due to pulsar transverse acceleration}  
   \maketitle

\section{Introduction}\label{sec:intro}

Pulsars are very stable rotators with rapid radio pulses. Pulsar timing analysis is based on the measurement of the precise pulse times of arrival (TOAs) at the telescope. The remarkable TOAs period stability opens up a wide range of potential applications, e.g. establish a pulsar-based time standard \citep{1994ApJ...428..713K,2012MNRAS.427.2780H}, detect nanohertz-frequency gravitational waves (GWs) \citep[e.g.][]{Jenet et al.(2004),Sesana et al.(2008),2009MNRAS.400..951V,Lee et al.(2011),Arzoumanian et al.(2015),Desvignes et al.(2016),Reardon et al.(2016),2019A&ARv..27....5B}, develop pulsar-based navigation techniques \citep{2006JGCD...29...49S,2013AdSpR..52.1602D,2019ApJS..244....1Z}, improve the solar system ephemeris \citep[e.g.][]{2010ApJ...720L.201C}, and study various scientific targets such as the interstellar medium and the solar wind \citep{2007MNRAS.378..493Y,2012MNRAS.422.1160Y}, etc. Using TEMPO\footnote{http://www.pulsar.princeton.edu/tempo} or TEMPO2\footnote{http://www.atnf.csiro.au/research/pulsar/tempo2} program, the TOAs can be transformed to the solar system barycenter which approximates an inertial frame. The spin period and period derivatives, the position, and the proper motion of the pulsar can also be obtained by fitting the TOAs to a pulsar timing model \citep{2006MNRAS.369..655H,2006MNRAS.372.1549E}. The difference between the observed and the model predicted TOAs are known as timing residuals. \par

Fitting errors in the parameters of the timing model will be revealed by systematic trends in the timing residuals. The patterns of these residuals due to various errors in the timing model, such as in pulsar spin period $P$ and its derivative $\dot{P}$, in positions, in proper motions have been well presented for decades in the literature \citep{2012puas.book.....L}. It was also noticed that the fitted period derivative may be different from its intrinsic value if the pulsar is accelerating in the gravitational field of the galaxy or a GC \citep{1992RSPTA.341...39P,1993ASPC...50..141P}. Some authors further studied the gravitational effects of dark matter halo on pulsar timing \citep{1990nep..conf..301S,1995A&A...297..607L,1996dsu..conf..252F,2007ApJ...659L..33S,2007MNRAS.382..879S,2011PhRvD..84d3511B,2014arXiv1405.1031B,2019arXiv191210210N}. Though the magnitudes of the timing parameters of the accelerating pulsar are widely concerned, the residual patterns for errors in the pulsar acceleration have not been reported previously. \cite{1992RSPTA.341...39P,1993ASPC...50..141P} showed that if a pulsar experiences a component of acceleration $a_{l}$ along the line of sight, the observed period derivative is generally related to the intrinsic value by $\dot P_{\rm obs}=\dot P_{\rm int}+(a_{l}/c)P $. Thus one may infer that its residual pattern should be identical with that due to an error in the spin period derivative $\dot{P}$. However, up to date, the residual pattern due to an error in the transverse acceleration of a pulsar is still unknown.

\cite{2006MNRAS.372.1549E} estimated the magnitude of the acceleration terms, for both the Galactic gravitational acceleration ($a\sim 10^{-11}~{\rm m~s^{-2}}$) and the greater acceleration in GCs ($a\sim 10^{-8}~{\rm m~s^{-2}}$), for all pulsars in the Australia Telescope National Facility (ATNF) catalogue \footnote{https://www.atnf.csiro.au/research/pulsar/psrcat/} \citep{2005AJ....129.1993M}. They showed that only radial acceleration term and Shklovskii term involving $a$ exceed 1 nanoseconds (ns) for a 20-yr observing campaign. Thus, all the other acceleration terms is neglected in TEMPO2 (see section 2.3 of Edwards et al. 2006). However, as we show later (in subsection \ref{the coupling effect}), the amplitudes due to the coupling effect of the pulsar transverse acceleration and the R$\rm{\ddot{o}}$mer delay, could be clearly larger than the maximum allowable systematic error of the TEMPO ($\sim 100$ ns) and TEMPO2 ($\sim 1$ ns) program, and may also be detectable for the pulsars in the nearest GCs, and thus needs more investigations.

In this work, we present a realistic simulation of the residual pattern for the pulsar with the transverse acceleration, and its possible applications are also discussed. The organization of this paper is as follows. The magnitude of the coupling effect and its influences on pulsar timing parameters are described in Section 2. The simulations and results are detailed in Section 3. Finally, we summarize the main results in Section 4.

\section{Methods}

\subsection{The coupling effect of the pulsar transverse acceleration and the R$\rm{\ddot{o}}$mer delay}
\label{the coupling effect}

The issue is essentially attributed to the geometric propagation delay \citep{2006MNRAS.372.1549E}. The displacement vector ($\boldsymbol{R}$) from the observatory to an isolated pulsar is the sum of the position of the pulsar ($\boldsymbol{R_{0}}$), the displacement of the pulsar ($\boldsymbol{k}$) in the time elapsed since the epoch $t_{\rm pos}$, and the barycentric position of the observatory ($\boldsymbol{r}$):

\begin{equation}\label{displacement vector}
\boldsymbol{R}=\boldsymbol{R_{0}}+\boldsymbol{k}-\boldsymbol{r}=\boldsymbol{R_{0}}+\boldsymbol{k}_{\parallel}-\boldsymbol{r}_{\parallel}+\boldsymbol{k}_{\perp}-\boldsymbol{r}_{\perp}.
\end{equation}
where the radial and transverse components are denoted by the subscripts, i.e. $i_{\parallel}=\boldsymbol{i}\cdot \boldsymbol{R_{0}}/|\boldsymbol{R_{0}}|$ and $\boldsymbol{i}_{\perp}=\boldsymbol{i}-i_{\parallel}\boldsymbol{R_{0}}/|\boldsymbol{R_{0}}|$, and $\boldsymbol{i}$ is an arbitrary vector. Neglecting terms of the order of $|\boldsymbol{R_{0}}|^{-3}$, the following relation is obtained:

\begin{equation}\label{displacement vector}
|\boldsymbol{R}|=|\boldsymbol{R_{0}}|+k_{\parallel}-r_{\parallel}+ \frac{1}{|\boldsymbol{R_{0}}|}(\frac{|\boldsymbol{k}_{\perp}|^2}{2}+\frac{|\boldsymbol{r}_{\perp}|^2}{2}-\boldsymbol{k}_{\perp}\cdot\boldsymbol{r}_{\perp})(1-\frac{k_{\parallel}}{|\boldsymbol{R_{0}}|}+\frac{r_{\parallel}}{|\boldsymbol{R_{0}}|}).
\end{equation}
The terms in the first pair of parentheses are the Shklovskii effect, annual parallax, and annual proper motion. The displacement $\boldsymbol{k}$ may be broken into the first and second derivatives \citep{2006MNRAS.372.1549E}:

\begin{equation}\label{veck}
\boldsymbol{k} = \boldsymbol{\mu} |\boldsymbol{R_{0}}|(t^{\rm psr}-t_{\rm pos}) +\frac{\boldsymbol{a}}{2}(t^{\rm psr}-t_{\rm pos})^2,
\end{equation}
where $\boldsymbol{\mu}$ is the velocity divided by the distance, $\boldsymbol{a}$ is the acceleration vector, and $t^{\rm psr}$ is the proper time measured at the pulsar.

\cite{2006MNRAS.372.1549E} assumed that $\boldsymbol{k} = \boldsymbol{\mu} |\boldsymbol{R_{0}}|(t^{\rm psr}-t_{\rm pos})$, and except for radial acceleration term and Shklovskii term, all the other acceleration terms is neglected in TEMPO2. Consequently, TEMPO2 does not include the coupling effect of the pulsar transverse acceleration and the R$\rm{\ddot{o}}$mer delay, i.e. for the annual proper motion term of equation \ref{displacement vector},

\begin{equation}\label{kdotr}
\frac{\boldsymbol{k}_{\perp}\cdot\boldsymbol{r}_{\perp}}{|\boldsymbol{R_{0}}|}=\boldsymbol{\mu}_{\perp}\cdot \boldsymbol{r}_{\perp} (t^{\rm psr}-t_{\rm pos})+ \frac{\boldsymbol{a}_{\perp}\cdot\boldsymbol{r}_{\perp}}{2 |\boldsymbol{R_{0}}|}(t^{\rm psr}-t_{\rm pos})^2,
\end{equation}
the second term of the right hand is neglected (see also equation 24 of Edwards et al. 2006). Thus, the magnitude of the timing residual of the coupling effect can be expressed as
\begin{equation}\label{Roemer delay}
\Delta_{{\rm R}\odot}^\prime = -\frac{\boldsymbol{a}_{\perp}\cdot\boldsymbol{r}_{\perp}}{2|\boldsymbol{R_{0}}| c}(t^{\rm psr}-t_{\rm pos})^2,
\end{equation}
where $c$ is the speed of light. If one takes $a_{\perp}\sim 5\times 10^{-8} {\rm m~s^{-2}}$, $R_{0} \sim 1 ~{\rm kpc}$, and $r_{\perp} \sim 1~{\rm AU}$ as an estimation for the case of a pulsar in a GC, the amplitude of the residual is about $160$ ns for a 20-yr observing campaign.

\subsection{The effects on pulsar timing residuals}

As shown above, the coupling effect of the pulsar transverse acceleration and the R$\rm{\ddot{o}}$mer delay does not include in the timing model of TEMPO2. Using the ecliptic coordinates ($\lambda , \beta$) and assuming a circular earth orbit that centered on the Sun, the R$\rm{\ddot{o}}$mer time delay is given by \citep{2012puas.book.....L}
\begin{equation}\label{Romer}
t_{\rm R}=t_{\oplus}\cos(\omega t-\lambda)\cos\beta,
\end{equation}
where $t_{\oplus}\simeq498.7$~s is the light travel time from the Sun to the Earth, $\omega\simeq2.0\times10^{-7}~{\rm s^{-1}}$ is the angular velocity of the Earth in its orbit, $\lambda$ and $\beta$ are the ecliptic longitude and latitude of the pulsar, and $\lambda=-\pi/2$ and $\beta=\pi/4$ are assumed in the following calculations.

A neglected transverse acceleration $a_{\rm \perp}$ will induce the errors in the coordinates, i.e. $\delta\lambda=[v_{\rm \perp}(t^{\rm psr}-t_{\rm pos})+\frac{1}{2}a_{\rm \perp} (t^{\rm psr}-t_{\rm pos})^2]\sin\xi/d_0$ and $\delta\beta=[v_{\rm \perp}(t^{\rm psr}-t_{\rm pos})+\frac{1}{2}a_{\rm \perp}(t^{\rm psr}-t_{\rm pos})^2]\cos\xi/d_0$, in which $v_{\rm \perp}$ is the transverse velocity, $\xi$ ($=\pi/4$ is taken in the following calculations) is the angle between the acceleration and the longitude line, and $d_0$ is the distance between the pulsar and the Earth. The errors give rise to periodic timing errors
\begin{equation}\label{RomerError}
\delta t_{\rm R}=t_{\oplus}\sin[\omega(t^{\rm psr}-t_{\rm pos})-\lambda]\cos\beta\delta\lambda-t_{\oplus}\cos[\omega (t^{\rm psr}-t_{\rm pos})-\lambda]\sin\beta\delta\beta.
\end{equation}
The corresponding error of a pulse phase is $\delta\phi=2\pi\nu \delta t_{\rm R}$, in which $\nu$ is the rotation frequency of the pulsar.

\section{Simulations}

\subsection{The residual pattern}

We firstly develop a phenomenological spin-down model to describe the pulse phase evolution, so that the model can be a tool for simulating the time-of-arrival (TOA) data. Assuming the pure magnetic dipole radiation in vacuum as the braking mechanism for a pulsar's spin-down \citep{2004hpa..book.....L}, the pulse phase evolution can be described as
\begin{equation}\label{dipole}
\ddot\phi(t) +\frac{8\pi^2B^2R^6\sin\theta^2}{3c^3I } \dot\phi(t)^3=0,
\end{equation}
where $B$ is the strength of the dipole magnetic field at the surface of the neutron star, $R~(\simeq10^6~{\rm cm})$, $I~(\simeq10^{45}~{\rm g~cm^2})$, and $\theta~(\simeq\pi/2)$ are the radius, moment of inertia, and angle of magnetic inclination, respectively. Solving equation (\ref{dipole}) and incorporating the coupling effect into the phase evolution, one can get the theoretical phase function $\Phi_{\rm T}(t)=\phi(t)+\delta\phi$.

\begin{figure}
\centering
\resizebox{12cm}{8cm}{\includegraphics{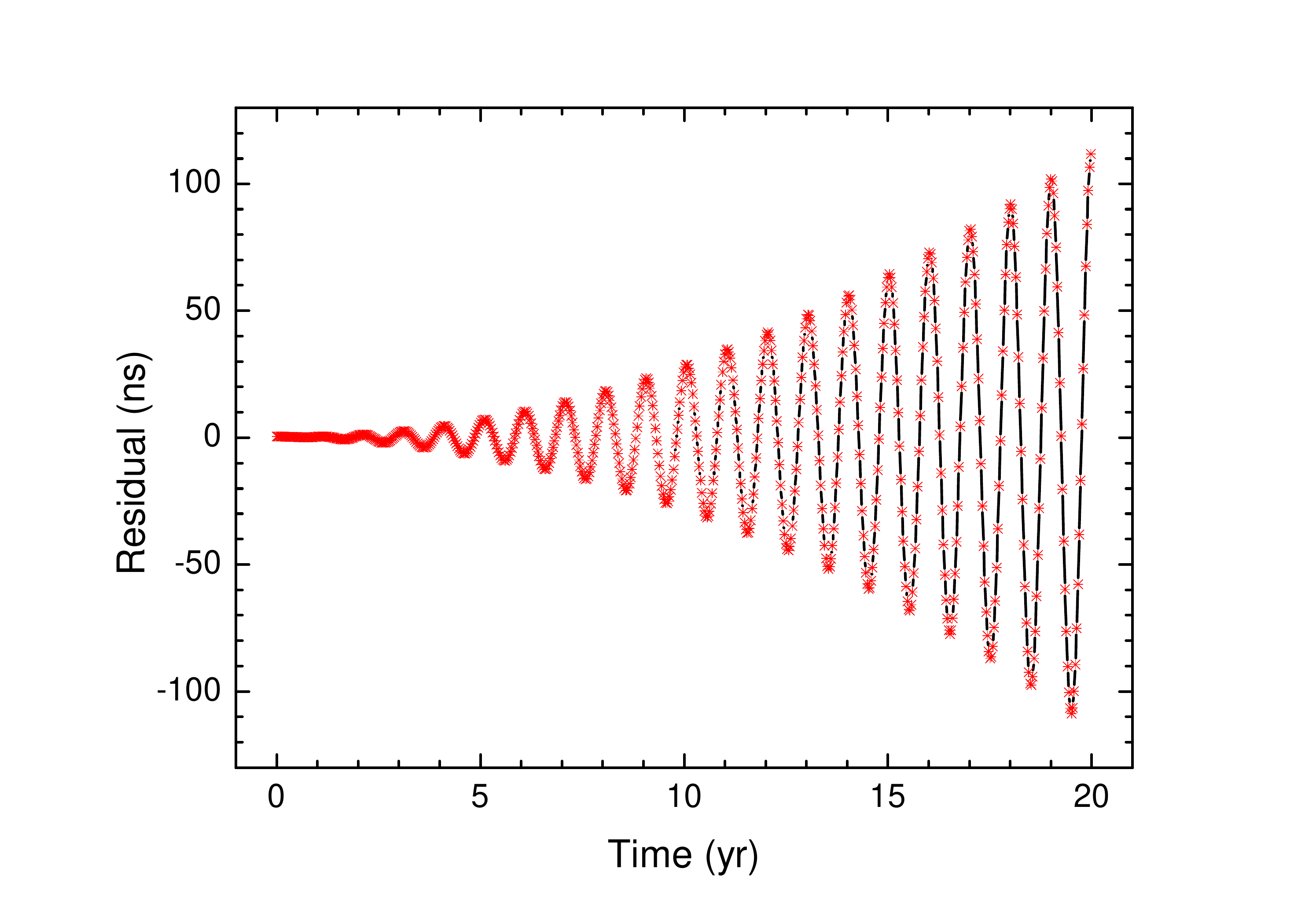}}
\caption {The simulated timing residual due to the coupling effect of the pulsar transverse acceleration and the R$\rm{\ddot{o}}$mer delay. The transverse acceleration $a_{\perp}= 5\times 10^{-8} {\rm m~s^{-2}}$ and the distance $d_{0}=1~{\rm kpc}$  are adopted in the simulations. }
\label{fig:1}
\end{figure}

We take the moderate initial values of $P_{0}=0.005~{\rm s}$ and $\dot P_{0}=1.0\times10^{-20}~{\rm s/s}$ for the MSP. The transverse acceleration $a_{\perp}= 5\times 10^{-8} {\rm m~s^{-2}}$ is assumed for the gravitational field of the GC \citep{1993ASPC...50..141P}. We assume a certain time interval $\Delta t_{\rm int}=10^6$~s between each two nearby TOAs to simulate the observed TOA series \citep{2015RAA....15..963X,2019ApJ...880..123X}. Since the rotational period is nearly constant, the pulsar spin frequency $\nu$, and its derivatives $\dot{\nu}$ and $\ddot{\nu}$ can be obtained by fitting the TOA series to the third order (cubic term) of its Taylor expansion over a time span $T_{\rm s}$,
\begin{equation}\label{phasefit}
\Phi(t) =\Phi_0 + \nu (t-t_0) + \frac{1}{2}\dot \nu (t-t_0)^2 + \frac{1}{6}\ddot\nu (t-t_0)^3.
\end{equation}
The pulsar period $P=1/\nu$ and its derivative $\dot P=-\dot \nu/\nu^2$ are quoted in place of $\nu$ and $\dot{\nu}$. Here we do not concern the relativistic frame transformation between observatory proper time and the pulsar proper time, thus $t=t^{\rm psr}$ and $t_0=t_{\rm pos}$ is simply taken in the simulation. The timing residual are conventionally defined as

\begin{equation}\label{residual}
({\rm Residual})\equiv\frac{\Phi_{\rm T}(t)-\Phi(t)}{2\pi\nu}.
\end{equation}
The results are shown in the upper right panel of Figure \ref{fig:1}. One can clearly see the residual with oscillation period of one year. \textit{The oscillation envelope of the residual due to velocity error are two straight lines, while two parabolic curves for the acceleration origin. This characteristic may be used to identify the transverse acceleration of a pulsar.} The root-mean-square (rms) of the residual due to the effect is about $100~{\rm ns}$, which is consistent with the estimate of equation \ref{Roemer delay}. Considering the larger distance $d_0$ for most of the GCs, the rms of practical residuals (inversely proportional to $d_0$) would be much smaller, and probably be drowned by other timing residuals. Hence the pattern of the residual has not been reported.

\begin{figure}
\centering
\resizebox{10cm}{7cm}{\includegraphics{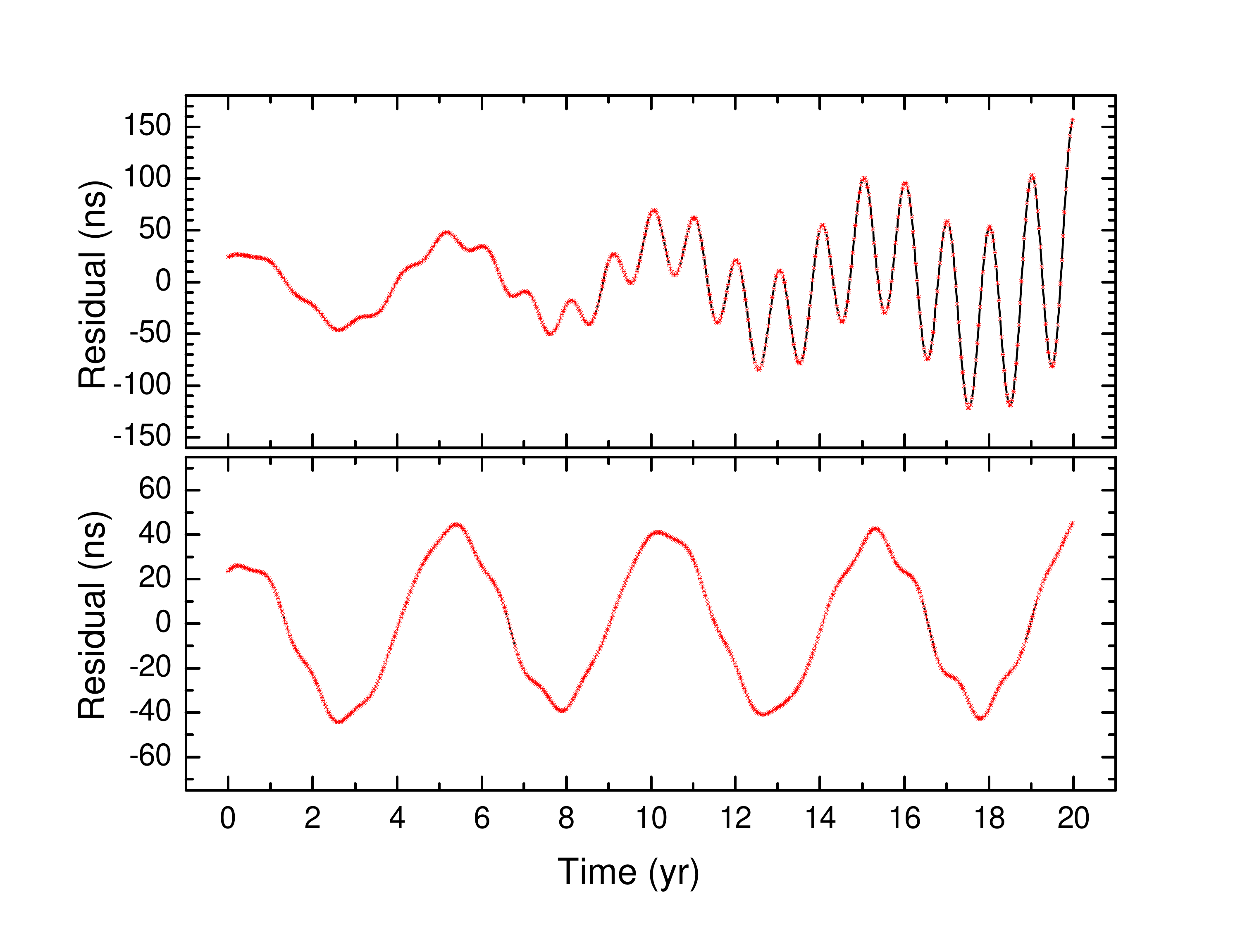}}
\caption {The simulated timing residuals with three oscillation modes and the transverse acceleration of a pulsar. Upper panel: the residual in fitting the phase with equation \ref{phasefit}.  Bottom panel: the residual in fitting the phase with equation \ref{phasefit2}.}
\label{fig:2}
\end{figure}

The pulsar transverse acceleration cannot be well determined if the oscillation drowned by other timing residuals. Based on the simulated envelopes of the residual curves, we speculate that the pulse phase with the annul effects may be characterized by

\begin{equation}\label{phasefit2}
\Phi(t) =\Phi_0 + \nu (t-t_0) + \frac{1}{2}\dot \nu (t-t_0)^2 + \frac{1}{6}\ddot\nu (t-t_0)^3+ \dot\nu' (t-t_0)^2\cos[\omega (t-t_0)+\lambda' ],
\end{equation}
in which $\dot\nu'$ and $\lambda'$ are free parameters. $\dot\nu'$ is responsible for the residual due to the coupling effect, and $\omega$ is the angular velocity of the Earth in its orbit. We perform again the residual simulation with the same parameters, but fit the phase sequences with equation \ref{phasefit2}. The magnitude of the simulated residual is fairly low ($\ll 1 ~{\rm ns}$), which means the effect can be well described with equation \ref{phasefit2}, at least for the case that the timing residuals are uncorrelated in pulsar timing.

If the annual effect correlates with ``red noises", which consist of low-frequency structures in timing residuals, it may cause some errors in estimating the parameters of the timing model and their uncertainties \citep{2011MNRAS.418..561C}. For this case, we firstly construct a phenomenological model for the pulsar timing residual with multi periodic oscillations, and the pulse phase evolution can be rewritten in the following form:
\begin{equation}\label{dipole1}
\ddot\phi +\frac{8\pi^2B^2R^6\sin\theta^2}{3c^3I }G(t)^{2} \dot\phi^3=0,
\end{equation}
in which $G(t)=1+\sum\limits_{i=1}^N k_i\sin(2\pi\frac{t}{T_i})$, and $k_i$ and $T_i$ is the magnitude and the period for the $i$-th oscillation, respectively. Then, we simulate the ``red noises" for three oscillation modes, and the period for two of them is about one year. Thus in the simulations, the three oscillation modes with $k_1=k_2=k_3= 10^{-4}$, $T_1=0.9$ yr, $T_2=1.1$ yr and $T_3=5.0$ yr are adopted \citep[see][for more details about the simulation]{2013IJMPD..2260012Z,2015RAA....15..963X}. For the annul effect, we also take $a_{\perp}=5\times 10^{-8}~{\rm m/s^2}$. The simulated residuals in fitting the phase with equation \ref{phasefit} and equation \ref{phasefit2} are shown in the upper panel and bottom panel of figure \ref{fig:2}, respectively. By comparing the two residuals, one can see that the annul effect of the transverse acceleration is fully separated out by equation \ref{phasefit2}, since the curve in bottom panel overlaps completely with the residual curve for the same oscillation modes but no the annul effect.

\begin{figure}
\centering
\resizebox{12cm}{9cm}{\includegraphics{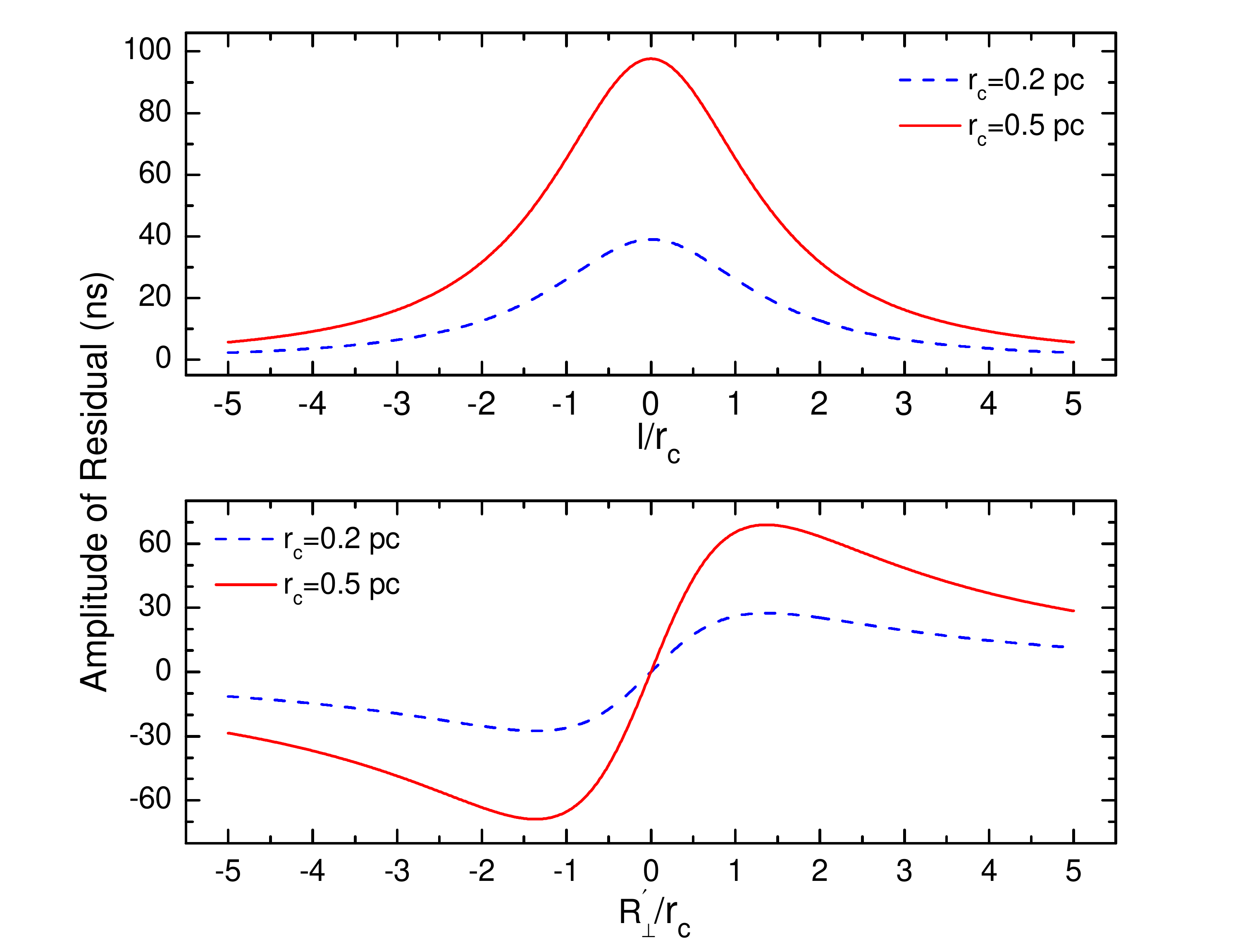}}
\caption {The amplitudes of timing residuals due to the coupling effect of the pulsar transverse acceleration and the R$\rm{\ddot{o}}$mer delay, for pulsars in a GC for 20-yr observing campaign. Upper panel: the residual amplitude with respect to $l$, and $R_{\perp}^{\prime}=r_{\rm c}$ is taken in the calculation. Bottom panel: the residual amplitude with respect to $R_{\perp}^{\prime}$, and $l=r_{\rm c}$ is assumed.}
\label{fig:3}
\end{figure}

\subsection{Residual amplitudes of pulsars in GCs}

In above simulations, $d_0=1~{\rm kpc}$ and $a_{\perp}=5\times 10^{-8}~{\rm m/s^2}$ are assumed. However in more practical cases, distances for many GCs are usually $\gtrsim 5~{\rm kpc}$, thus inducing residual with amplitude of only tens nanoseconds, which is relatively small, even though for these pulsars with highest timing precision \citep[e.g., ][]{2019MNRAS.490.4666P}. However, a pulsar's transverse acceleration due to the GC's mean field differs with its location with respect to the center of gravity (CoG) and the GC geometry.

We define the plane running through the CoG and perpendicular to our line-of-sight as $O$, the core radius of the GC as $r_{\rm c}$, the impact parameter for each pulsar from the CoG as $R_{\perp}^{\prime}$, and the line-of-sight position going perpendicularly through $O$ as $l$, and the pulsar's spherical radius $r^{\prime}$ ($=\sqrt{R_{\perp}^{\prime 2}+l^2}$). Then the cluster acceleration felt at any given radius out from the core can be written as \citep{2017ApJ...845..148P}
\begin{equation}\label{field}
a_{\rm r}(r^{\prime})=-4\pi G\rho_{\rm c}r_{\rm c}^3r^{{\prime}-2}[\sinh^{-1}(\frac{r^{\prime}}{r_{\rm c}})-\frac{r^{\prime}}{r_{\rm c}\sqrt{1+(r^{\prime}/r_{\rm c})^2}}]
\end{equation}
where $a_{\rm r}(r^{\prime})$ is the mean-field acceleration, and $\rho_{\rm c}$ is the core density. The King density profile which most strongly sets the GC potential \citep{1962AJ.....67..471K},
\begin{equation}\label{King Model}
\rho(r^{\prime})\simeq\rho_{\rm c}[1+(r^{\prime}/r_{\rm c})^2]^{-\frac{3}{2}},
\end{equation}
is included in the model. We get the transverse acceleration $a_{\perp}$ by projecting the acceleration $a_{\rm r}(r^{\prime})$ along the transverse direction by a factor of $R_{\perp}^{\prime}/r^{\prime}$. Substituting typical values of $\rho_{\rm c}=10^6~M_{\bigodot} {\rm pc}^{-3}$ and $d_0=5~{\rm kpc}$ for the GC, and using equation \ref{Roemer delay}, we obtain the amplitude of the residuals for various $l$ or $R_{\perp}^{\prime}$, as shown in figure \ref{fig:3}. As expected, the amplitudes are very sensitive to $r_{\rm c}$, and actually proportional to the mass of the core. The results also imply that the residuals due to the coupling effect may probably be observed, particularly for those pulsars, which distribute in the vicinity of big cores and locate near the $O$ plane.

The amplitudes of the residuals presented in Figure \ref{fig:1} and Figure \ref{fig:3} do not depend on the initial values of pulsar spin parameters (e.g. $P$ and $\dot P$ ), and thus are universal for different pulsars. The simulations for MSPs or normal pulsars in the galaxy with $g_{\rm G}=10^{-10}~{\rm m/s^2}$ are also performed, and we found that the acceleration is small and have little influence on pulsar timing.

\section{Summary}
We simulated the timing residuals of a pulsar due to the coupling effect of the transverse acceleration and the R$\rm{\ddot{o}}$mer delay, for a 20-yr observing campaign. It is found that the envelopes of the residuals due to the pulsar acceleration are two parabolic curves, which may be used to identify the transverse acceleration of a pulsar. Even if drowned by other timing residuals, they can also be well modeled and separated out with equation \ref{phasefit2}. However, the annual effect on pulsar timing is usually relatively small: for pulsars in the galactic field, the acceleration due to the galactic potential is of the order of about $g_{\rm G}=10^{-10}~{\rm m/s^2}$, which inducing timing residual $< 1 ~\rm{ns}$. Only for these pulsars in GCs, this effect is possibly needed. We calculated the amplitudes for pulsars distribute around the cores of GCs, and the results imply that the coupling effect may probably be observed for pulsars locating near the $O$ plane in close clusters. We expect to gain more details of the timing residuals and much deeper understanding of GC dynamics using future larger samples of MSPs with higher precision data, to be brought by China's Five-hundred-meter Aperture Spherical radio-Telescope (FAST) and the future Square Kilometre Array (SKA).

\begin{acknowledgements}
We thank Dr. Jian-ping Yuan for valuable discussions. We thank the anonymous referee for comments and suggestions that led to a significant improvement in this manuscript. This work is supported by National Natural Science Foundation of China under grant Nos. 11803009 and 11603009, and by the Natural Science Foundation of Fujian Province under grant Nos. 2018J05006, 2018J01416 and 2016J05013.

\end{acknowledgements}

\end{document}